\documentclass{aastex62}
\usepackage[utf8]{inputenc}

\shorttitle{Detrending timeseries data of strictly periodic astrophysical objects}
\shortauthors{Pr\v sa, Zhang \& Wells}

\newcommand{\rev}[1]{#1}

\begin{document}

\title{FINDING \rev{THE} NEEDLE IN \rev{A} HAYSTACK: DETRENDING PHOTOMETRIC TIMESERIES DATA OF STRICTLY PERIODIC ASTROPHYSICAL OBJECTS}

\author[0000-0002-1913-0281]{Andrej Pr\v sa}
\affiliation{Villanova University, Dept.~of Astrophysics and Planetary Science, 800 Lancaster Ave, Villanova PA 19085}

\author{Moses Zhang}
\affiliation{Byram Hills High School, 10 Tripp Lane, Armonk NY 10504}
\affiliation{Villanova University, Dept.~of Astrophysics and Planetary Science, 800 Lancaster Ave, Villanova PA 19085}

\author{Mark Wells}
\affiliation{Villanova University, Dept.~of Astrophysics and Planetary Science, 800 Lancaster Ave, Villanova PA 19085}
\affiliation{Penn State University, Department of Astronomy and Astrophysics, Eberly College of Science, University Park, PA 16802}

\correspondingauthor{Andrej Pr\v sa}
\email{aprsa@villanova.edu}

\begin{abstract}
    Light curves of astrophysical objects \rev{frequently contain strictly periodic signals}. In those cases we can use that property to aid the detrending algorithm to fully disentangle an \emph{unknown} periodic signal and an \emph{unknown} baseline signal with no power at that period. The periodic signal is modeled as a discrete probability distribution function (pdf), while the baseline signal is modeled as a residual timeseries. \rev{Those two components are disentangled} by minimizing the length of the residual timeseries w.r.t.~the per-bin pdf fluxes. We demonstrate the use of the algorithm on a synthetic case, on the eclipsing binary KIC 3953981 and on the eccentric ellipsoidal variable KIC 3547874. We further discuss the parameters and the limitations of the algorithm and speculate on the two most common use cases: detrending the periodic signal of interest and measuring the dependence of instrumental response on controlled instrumental variables. \rev{A more sophisticated version of the algorithm is released as open source on github and available via pip.}
\end{abstract}

\keywords{methods: numerical, methods: data analysis, binaries: eclipsing, stars: variables: general, stars: individual (KIC 3953981, KIC 3547874), planets and satellites: detection}

\section{Introduction}

Photometric timeseries data are one of the fundamental observational pillars in astronomy. Interpreting these data relies critically on our ability to distinguish between the intrinsic part of the observed signal (the one that corresponds to the object of interest) and the extrinsic part (detector response, optical characteristics of the instrument, atmospheric seeing, interstellar features, \rev{etc.}). This interplay makes the observed timeseries often exceedingly complex in structure and overwhelmingly difficult to interpret. To perform high precision, high fidelity modeling of objects of interest, we need to resort to one of the following strategies: (1) make a number of simplifying assumptions that keep the model tractable; (2) include the extrinsic part of the signal in the model by introducing heuristic and/or empirical description such as, for example, Gaussian processes \citep{foremanmackey2017}; or (3) remove the extrinsic part altogether. No matter the approach, there is an omnipresent danger of attributing some of the intrinsic signal to extrinsic sources, or vice versa, as we rarely know all the sources that contribute to and all the signatures that are found in the data to a sufficient accuracy to model them adequately.

In recent years\rev{,} significant effort was invested in developing the methods for removing systematic artifacts from the \textsl{Kepler} data \citep{borucki2010}. The \textsl{Kepler} pipeline incorporates Pre-search Data Conditioning (PDC; \citealt{stumpe2012, smith2012}), within which the systematics are removed with a combination of data detrending and cotrending against weighted cotrending basis vectors derived from the time series structure most-common to all neighbors of the scientific target \citep{thompson2016}. Another detrending approach based on cotrending basis vectors is implemented within the Variational Bayes framework \citep{aigrain2017}; it can correct for discontinuities and bulk trends. The \textsl{Kepler} Guest Observer Office developed the PyKE tool for IRAF \citep{still2012}, with the \texttt{kepdetrend} and \texttt{kepcotrend} tasks for detrending the data using polynomial fits. \citet{prsa2011} used self-adjusting polynomial chains to remove trends from eclipsing binary light curves that are collected in the \textsl{Kepler} Eclipsing Binary catalog \citep{kirk2016}. An algorithm devised to detrend exoplanetary light curves based on coherent median filtering, \textsl{phasma} \citep{jansen2018}, was used to detrend 165 \textsl{Kepler} light curves. A slew of methods, either general \citep{vandenburg2014} or dedicated for a particular purpose\footnote{A comprehensive list is maintained at \texttt{https://github.com/KeplerGO/lightkurve} by the \textsl{Kepler}/K2 Guest Observer office.} have been developed for K2 \citep{howell2014} as well, and their applications and further advancements are ongoing for space surveys such as TESS \citep{ricker2015}. Utilizing and developing methods for robust, reliable detrending of the timeseries data has thus been at the forefront of modern research.

In this paper we present a detrending algorithm that targets strictly periodic variables. We make no assumptions about the functional form of the periodic signal, nor do we place any constraints on the extraneous signal. By ``extraneous'' we imply \emph{all} parts of the overall signal that are not strictly periodic, including those of an astrophysical origin. The only parameter to the algorithm is the period itself; the output is a fully disentangled periodic component of the signal and the remaining, extraneous timeseries. In Section \ref{sec:algorithm} we provide a mathematical description of the method; in Section \ref{sec:demo} we present three case studies: a synthetic signal, an eclipsing binary star and an eccentric ellipsoidal variable, also known as a \emph{heartbeat} star; in Section \ref{sec:limitations} we discuss some inherent limitations of the algorithm and in Section \ref{sec:conclusion} we provide a general discussion and conclusions.

\section{The Algorithm} \label{sec:algorithm}

Consider a signal that can be described as:
\begin{equation} \label{eq:signal}
    S(t) = X(t) + Y(t) + \varepsilon(t),
\end{equation}
where $X(t)$ is a strictly periodic signal with period $P$, $Y(t)$ is a ``baseline'' signal with no power at period $P$, and $\varepsilon(t)$ is a purely stochastic component. We will refer to signal $X(t)$ as \emph{synchronous} and signal $Y(t)$ as \emph{asynchronous}. Eq.~(\ref{eq:signal}) can then be written as:
\begin{equation} \label{eq:signal_phase}
    S(t) = X(E P \Phi(t)) + Y(t) + \varepsilon(t),
\end{equation}
where $\Phi(t) = [(t-t_\mathrm{ref}) \, \mathrm{mod} \, P] / P$ is phase on the $[0, 1)$ interval, $t_\mathrm{ref}$ is a reference time (i.e.~a phase shift), and $E$ is an integer that enumerates consecutive cycles. Let us further define the discretized counterparts of the above quantities: with $t_k$ we denote the timestamp integrated over $\Delta t_k \equiv t_{k+1/2}-t_{k-1/2}$; with $S_k \equiv S(t_k)$ the discretized signal at timestamp $t_k$; with $X_k \equiv X(\Phi_k) \equiv X(\Phi(t_k))$ the discretized synchronous component at timestamp $t_k$; with $Y_k \equiv Y(t_k)$ the discretized asynchronous component at timestamp $t_k$; and with $\varepsilon_k \equiv \varepsilon(t_k)$ the discretized stochastic component at timestamp $t_k$. Note that the discretized phase, $\Phi_k$, has its own scale that depends on $t_k$ \emph{and} the period $P$, so the sequence $\{\Phi_k\}$ will not be sorted as $t_k$ because of the modulo operator. \rev{As our goal is to disentangle the synchronous and the asynchronous components of the signal, we bundle the asynchronous and the stochastic components into one variable, $\widetilde Y_k \equiv Y_k + \varepsilon_k$.} A discretized version of Eq.~(\ref{eq:signal_phase}) can then be written as:
\begin{equation}
    S_k = X(E P \Phi_k) + \widetilde Y_k.
\end{equation}
Given an observed timeseries $O_k \equiv O(t_k)$, we note that $S_k = O_k$, i.e.~that the \emph{entire} observed signal is explained by the synchronous component $X(\Phi_k)$, asynchronous component $Y_k$ and stochastic component $\varepsilon_k$. A typical situation in astronomical observations is that $O_k$ is known, $P$ is determined to a reasonable accuracy by phase-folding, i.e.~by using Phase Dispersion Minimization (PDM; \citealt{stellingwerf1978}), Analysis of Variance (AoV; \citealt{schwarzenberg1989}) or Box Least Squares (BLS; \citealt{kovacs2002}), and $X(\Phi_k)$ and $\widetilde Y_k$ are unknown. Commonly, $\widetilde Y_k$ will either be assumed stochastic ($Y_k = 0$, $\widetilde Y_k \equiv \varepsilon_k$) or be removed from the data by detrending ($Y_k = f(t_k)$ for some detrending function $f$), and $X(\Phi_k)$ will be fit by a physical model of interest. 

The synchronous component $X(\Phi_k)$ can be represented by a discrete probability density function (pdf) that describes the unknown strictly periodic signal. This means that we make no assumptions about the analytical or empirical form of the signal, just that it can be adequately represented by a discrete pdf.

The asynchronous component $\widetilde Y_k$ can be represented by the residual timeseries:
\begin{equation} \label{eq:async}
    \widetilde Y_k = O_k - X(E P \Phi_k).
\end{equation}
Next, we define the length operator on a generic discrete function $f_j \equiv f(t_j)$ as:
\begin{equation}
    \hat L f = \sum_j \left[ (f_{j+1}-f_j)^2 + (t_{j+1}-t_j)^2 \right]^{1/2}\,,
\end{equation}
where index $j$ runs over all $k$. From Eq.~(\ref{eq:async}) it follows that $X(\Phi_k)$ will be an optimal discrete pdf for the synchronous component of $O_k$ when $\hat L \widetilde Y_k = \mathrm{min}$. In other words, $X(\Phi_k)$ and $\widetilde Y_k$ will be optimally disentangled when the length of the $\{\widetilde Y_k\}$ sequence is minimal. As each $X(\Phi_k)$ pdf has a corresponding length $\hat L \widetilde Y_k$, minimizing $\hat L \widetilde Y_k$ optimizes the $X(\Phi_k)$ pdf.

To achieve this, the $X(\Phi_k)$ pdf is iteratively adjusted. Barring any obvious non-stationary trends, drifts or unit-root processes in the generative function for $\widetilde Y_k$, a reasonable starting point for the $X(\Phi_k)$ pdf is a phase-folded average of $S_k$. Otherwise, some detrending might be necessary to make the data stationary and then assign a phase-folded average to the initial $X(\Phi_k)$ pdf. Each ``bin'' $i$ of the discrete pdf affects the length of $\widetilde Y_k$ proportionally to $\partial \hat L \widetilde Y_k / \partial X_i(\Phi_k)$, so we need to take a differential step along that gradient until $\hat L \widetilde Y_k$ is minimized. Differential steps can be computed either by differential corrections (DC; \citealt{euler1755}) or by any robust heuristic methods (simulated annealing, conjugate gradients, etc.\rev{ --- see \citet{press2007} for further details}). 

\bigskip

A simple form of the algorithm proceeds as follows. 
\begin{description}
\item[Phase-fold the observed light curve] the composite signal $S(t)$ is optionally shifted using the time shift parameter $t_\mathrm{ref}$ (i.e.,~if we desire any particular feature in a light curve to land at any particular phase) and phased on the period of interest, $P$. This generates the undetrended phased light curve.

\item[Generate a discrete pdf for the synchronous signal] depending on the cadence of observations and on the period, we assign a number of bins $N_b$ to the discrete pdf that will describe the synchronous signal in phase space. Each pdf bin must have at least one observation. The choice for the initial pdf affects the number of iterations the algorithm will require to minimize the length of the residual curve; barring light curves that are severely dominated by the asynchronous component, a good choice for the initial pdf is the per-bin averaged phase-folded observed light curve.

\item[Compute the asynchronous curve] given observations $O_k(t_k)$ and the discrete pdf $X_k(\Phi_i)$, unfold $X_k(\Phi_i)$ to $X_k(t_j)$ for each timestamp $t_j$ in the observed dataset. As $X_k(\Phi_i)$ is discrete, linearly interpolate $X_k(t_j)$ from the two bounding bin ranges to minimize numerical artifacts. Then compute the residual curve $\widetilde Y_k \equiv O_k-X_k$.

\item[Compute partial derivatives] the slopes in the parameter space are described by the partial derivatives $\partial \hat L \widetilde Y_k/\partial X_i(\Phi_k)$, where $k$ enumerates iterations and $i$ enumerates pdf bins. These derivatives inform the algorithm how the length of the residual curve is affected by the change in each pdf bin value. We use the symmetrically discretized differences to approximate these derivatives:
\begin{equation} \label{eq:slopes}
    \frac{\partial \hat L \widetilde Y_k}{\partial X_k(\Phi_i)} \to \frac{\hat L \widetilde Y_k(\cdots X_{k,i} + \frac 12 \delta X_{k,i} \cdots) - \hat L \widetilde Y_k(\cdots X_{k,i} - \frac 12 \delta X_{k,i} \cdots)}{\delta X_{k,i}},
\end{equation}
where $\delta X_{k,i}$ is the change in the $i$-th pdf bin value. 

\item[Compute an improved pdf] each per-bin pdf value is changed along the negative gradient computed in the previous step:
\begin{equation}
    X_k (\Phi_i) \to X_k(\Phi_i) - \xi \frac{\partial \hat L \widetilde Y_k}{\partial X_k(\Phi_i)},
\end{equation}
where $\xi$ is a constant that determines the step size along the slope. 

\item[Iterate until the pdf is optimized] each change in the pdf reduces the length of the residual asynchronous curve. We repeat the process of modifying the pdf and computing the asynchronous curve length until the length in two consecutive steps changes by less than a prescribed tolerance $\eta$.

\end{description}

The result of this process is the \rev{separation of the strictly periodic pdf signal at period $P$ from} ``everything else\rev{,}'' i.e.~the asynchronous and the stochastic components of the net signal. This simple version of the algorithm depends on 4 parameters: number of pdf bins $N_b$, pdf per-bin value variation $\delta X_{k,i}$, descent step size $\xi$, and convergence stop criterion $\eta$. We provide some considerations on choosing the values of these parameters in Section \ref{sec:limitations} but stress that a more robust version of this algorithm could easily optimize these values as part of the execution process. \rev{We propose some efficient optimization steps in Section \ref{sec:conclusion}.}

\section{Case Studies} \label{sec:demo}

\subsection{Synthetic case}

\begin{figure}[t!]
    \centering
    \includegraphics[width=0.85\textwidth]{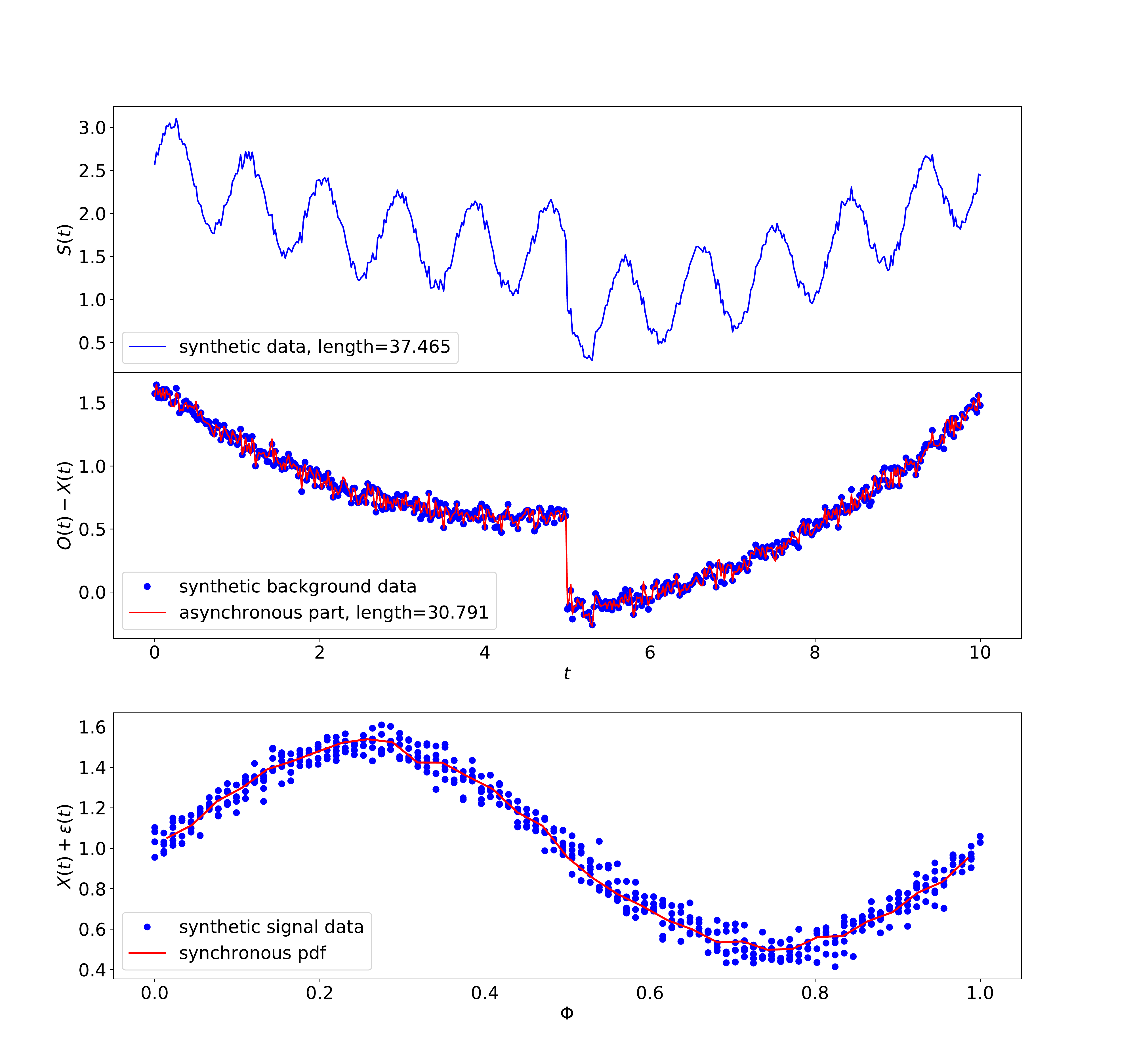} \\
    \caption{\textbf{Top}: synthetic (mock) signal that consists of a strictly periodic component with period $P=1$ unit, a non-periodic baseline that consists of a quadratic and an exponential function and features a discrete jump, and a stochastic noise component, cf.~Eq.~(\ref{eq:synthetic}). \textbf{Middle}: minimum length residual curve with the strictly periodic signal subtracted out. \textbf{Bottom}: strictly periodic component of the signal determined by minimizing the length of the residual curve.}
    \label{fig:synthetic}
\end{figure}

To demonstrate and test the algorithm performance, we start with mock data. The composite signal $S(t)$ is generated using the following prescription:
\begin{equation} \label{eq:synthetic}
    S(t) = \left[ 1 + \frac 12 \sin \left( 2\pi \frac{t-t_0}{P} \right) + \frac 1{20} (t-5)^2 + e^{t/10-1} + \mathcal N(0, 0.05) \right] - 0.75 \mathcal H(5),
\end{equation}
where $X(t) \equiv 1+\sin(2\pi (t-t_0)/P)/2$ is the strictly periodic term with period $P$ and time offset $t_0$, the term $(t-5)^2/20$ is a quadratic part of the asynchronous component, $e^{t/10-1}$ is an exponential part of the asynchronous component, $\mathcal N(0, 0.05)$ is stochastic noise, and $\mathcal H(5)$ is the Heaviside step function that equals 0 for $t < 5$ and 1 for $t \geq 5$.
We generate S(t) for the time span of $10$ units, $P=0.91$ units and $t_0=0$ units. Thus, $S(t)$ is a composite function that consists of a strictly periodic signal, of a non-periodic asynchronous component with a discrete jump, and of a stochastic noise component. The top panel in Fig.~\ref{fig:synthetic} depicts a realization of this generative function.

The algorithm was then run on these data with $N_b = 33$, $\delta X_{k,i} = 5 \times 10^{-4}$ units, $\xi = 10^{-3}$ and $\eta = 10^{-8}$. The results are depicted in the middle panel (asynchronous component) and in the lower panel (synchronous component). As this is a mock example, it gives us a benefit of comparing it against the known input. The synthetic data are depicted in blue, and the disentangled signal is depicted in red. Running the algorithm numerous times shows that, on average, the signal pdf is reproduced within the data uncertainty every time. This serves as proof of concept that the algorithm performs as expected.

\subsection{Eclipsing Binary Star KIC 3953981}

Astronomical objects that lend themselves readily to this algorithm's principal use are eclipsing binary stars (EBs). EBs serve as ideal astrophysical laboratories to measure their components' masses and radii: the well-understood laws of two-body dynamics and geometric considerations that lead to eclipses render inversion (the process to determine parameters from light and radial velocity curves) a tractable problem. As these masses and radii are used to calibrate the mass-radius relationship and proliferate through the rest of stellar and galactic astrophysics, as well as serve as principle distance gauges, particular effort needs to be invested to remove any and all artifacts from the observed data.

A majority of EB light curves are strictly periodic. A small number of EBs exhibit eclipse timing variations (ETVs) of the order of seconds to minutes, rarely hours. For as long as the bin size is larger than the (double) amplitude of the ETV, such light curves can also be considered strictly periodic. Any signals that are not commensurate with the orbital period (pulsations, non-synchronous spots, differential rotation effects, etc), and any dynamically variable features (transient spots, outbursts, etc) will be removed from the synchronous pdf by the algorithm. In some cases that is an added benefit as the modeling of the ``blemish''-free light curve might be favored over a complex model that includes dynamically and temporally variable effects.

\begin{figure}[t!]
    \centering
    \includegraphics[width=0.49\textwidth]{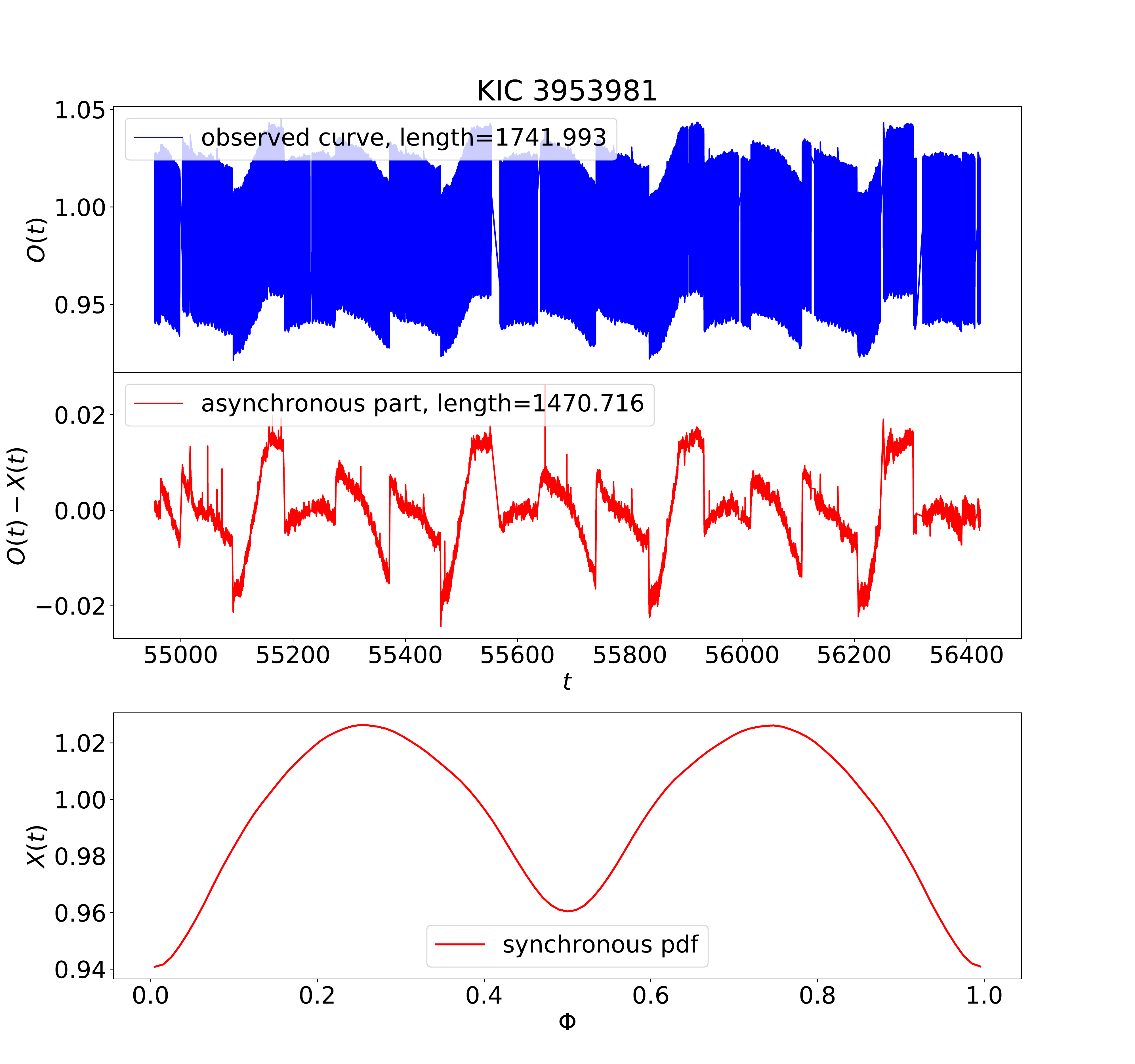}
    \includegraphics[width=0.49\textwidth]{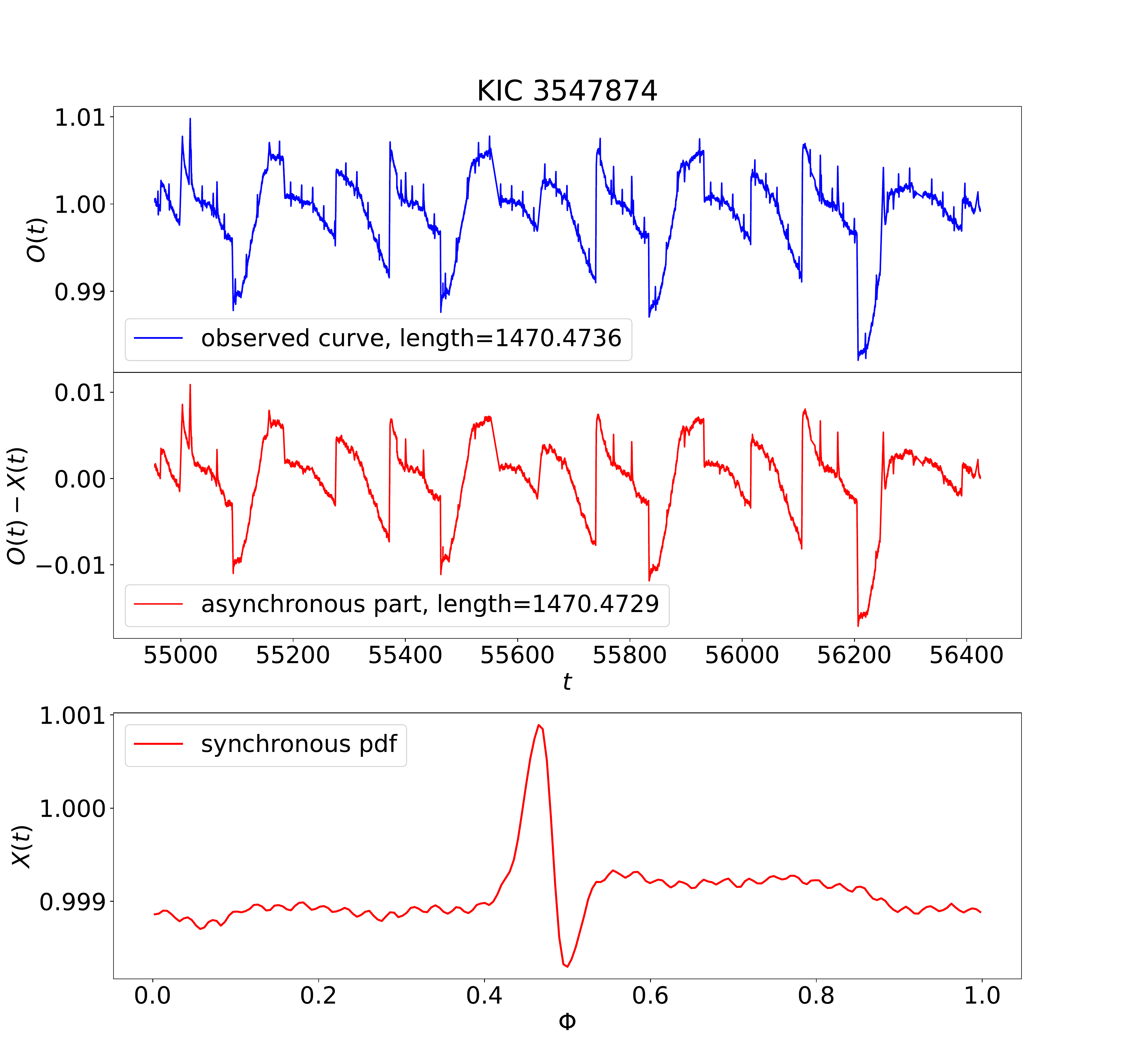} \\
    \caption{
    \textbf{Top}: simple aperture photometry (SAP) light curve of KIC 3953981, an eclipsing binary star (left) and KIC 3547874, an eccentric ellipsoidal variable (right) observed by \textsl{Kepler}.  \textbf{Middle}: minimum length asynchronous part of the observed curve with the strictly periodic pdf subtracted out. \textbf{Bottom}: synchronous pdf determined by minimizing the length of the asynchronous curve.
    }
    \label{fig:observed}
\end{figure}

We ran the algorithm on KIC 3953981, a circular, short-period eclipsing binary with orbital period $P=0.492$ days and $\mathrm{BJD}_0 \equiv t_\mathrm{ref}=54953.8225$, observed by the \textsl{Kepler} spacecraft over the course of $\sim$4 years. We used simple aperture photometry (SAP) data -- a \textsl{Kepler} pipeline deliverable that 
does \emph{not} remove any trends. Given the light curve's short orbital period and smooth light curve variability, a modest number of bins, $N_b=101$, was sufficient to describe the synchronous pdf (cf.~Fig.~\ref{fig:observed}, left).
Other parameters of the algorithm were: $\delta x_{k,i} = 5 \times 10^{-4}$ normalized flux units, $\xi = 10^{-5}$ and $\eta = 10^{-8}$. Despite the demonstrably complex baseline, the algorithm performed convincingly well.

\newpage

\subsection{Eccentric Ellipsoidal Variable KIC 3547874}

Another of \textsl{Kepler} mission's notable successes was the observation of over a hundred eccentric ellipsoidal variables \citep{thompson2012}; their light curve resemblance to electro-cardiograms earned them the name of \emph{heartbeat} stars. The components in these highly eccentric binaries become highly distorted during their periastron fly-by due to the increase in tidal interaction, causing a spike in flux due to an increase in cross-section and reflection. In addition, periodic periastron ``pinches'' by tides can induce pulsations in one or both components that are commensurate with the orbital period and are revealed by the short sinusoidal oscillations in the light curve. Because of this, heartbeat stars are ideal testbeds to study tidally excited pulsations and resonance locking \citep{fuller2017, hambleton2018}.

Much like EBs, the orbital periods of hearbeat stars are largely constant, thus lending themselves readily to the algorithm. An added challenge is that, unlike EBs, heartbeat star light curves are usually very low in amplitude, and the signal is buried in noise. This makes detrending all the more difficult using conventional detrending algorithms.

We ran the algorithm on KIC 3547874, an eccentric ellipsoidal variable with orbital period $P=19.692$ days and $\mathrm{BJD}_0 \equiv t_\mathrm{ref} = 54989.421$, observed by the \textsl{Kepler} spacecraft over the course of $\sim$4 years. As before, we used the SAP data as input to the algorithm. Even though tidally induced pulsations were not evident in the original data, we used $N_b=201$ bins to make sure we have sufficient resolution to detect them (cf.~Fig.~2, right). Indeed, as the bottom panel demonstrates, the algorithm successfully extracted the signature of tidally induced pulsations at the 37$^\mathrm{th}$ harmonic of the orbital period. Other parameters of the algorithm were: $\delta x_{k,i} = 5 \times 10^{-4}$ normalized flux units, $\xi = 10^{-4}$ and $\eta = 10^{-9}$. Even though the peak-to-peak amplitude of the periastron brightening is only $\sim 0.3\%$ and the peak-to-peak amplitude of tidally induced pulsations is $\sim 0.01\%$, the algorithm was able to recover their signatures. Provided that the peak-to-peak amplitude of the background light curve is $> 2\%$, we conclude that the algorithm, despite its simplicity, performed better than expected.

\section{Limitations of the Algorithm} \label{sec:limitations}

The simple form of the algorithm presented here depends on the input light curve, the orbital period, and 4 parameters: $N_b$, $\delta X_{k,i}$, $\xi$ and $\eta$. All 4 parameters can in principle be hidden through automation, but we felt that doing so will obfuscate the clarity of the principle behind the algorithm and have opted to defer such optimization to subsequent code development. Here we discuss the roles of these parameters in more detail, along with possible optimization routes.

\begin{description}
\item[Number of bins $N_b$] measures the smoothness of the discrete pdf. If the number is too small, the pdf will not adequately capture the features present in the data; if the number is too large, the computation time cost will be prolonged and numerical errors due to low number statistics per bin \rev{may become significant}. Each additional bin adds a computation of one extra partial derivative $\partial \hat L \widetilde Y_k/\partial X_k(\Phi_i)$ and adds one extra degree of freedom. As the feature space is rarely uniform, the choice of an equidistant pdf as used above is likely inferior to advanced methods of binning data \rev{(see \citet{hogg2008} for an example)}. The algorithm in its present form also does not allow for empty pdf bins, so a reasonable value for $N_b$ has to be smaller than the number of observations per orbital period, possibly significantly smaller if a non-uniform cadence is used.

\item[Finite difference $\delta X_{k,i}$] used for the calculation of slopes in the parameter space. If $\delta X_{k,i}$ is too small, the algorithm will suffer from numerical noise. If too large, the slope will be poorly approximated and, depending on the topology of the parameter space, convergence might be compromised. The value of $\delta X_{k,i}$ is theoretically driven by the degree of linearity of the space of partial derivatives $\partial \hat L Y/\partial X_{k,i}$: $\delta X_{k,i}$ should be the same order of magnitude as the extent at which $\partial \hat L Y/\partial X_{k,i}$ is largely constant. Moreover, we adopt a single parameter for all $N_b$ finite differences $\delta X_{k,i}$, $i=1 \dots N_b$; for the same reasons as above, this can be optimized further by considering the feature properties across the phase space.

\item[Step size $\xi$] provides the extent of per-bin change of the pdf. If $\xi$ is too small, the algorithm will converge slowly; if too large, the algorithm will start to oscillate about the minimum and become unstable. This is also the most sensitive of all algorithm parameters and not trivial to estimate in advance. The telltale sign of the value of $\xi$ being too large is the oscillatory behavior of pdf corrections. The optimization process would thus identify the largest value of $\xi$ for which no oscillatory behavior is exhibited. This will vary from case to case, depending on the amplitude of the synchronous signal, the signal-to-noise ratio, and on the topology of the underlying parameter space. 

\item[Convergence criterion $\eta$] determines the state at which the algorithm considers the signal pdf and the asynchronous component disentangled. If $\eta$ is too large, the two components will not be adequately separated; if too small, the number of iterations will increase to the point of inefficiency due to a high time cost, or even fail to converge if numerical artifacts prevent the difference in consecutive lengths $\hat L Y_{k+1}$ and $\hat L \widetilde Y_k$ \rev{from dipping below} $\eta$. Convergence is asymptotic and optimizing $\eta$ would imply finding a good trade-off between the accuracy of disentangling and the computational time cost based on the analysis of the asymptotic convergence behavior.

\end{description}

Another potential deficiency inherent to the simple version of the algorithm stems from using linear interpolation to compute intra-bin values of \rev{the} pdf. A more general solution would be to employ a higher order univariate spline. This issue diminishes with a sufficiently large bin number $N_b$; the telltale sign of the issue is a residual periodic signal in the asynchronous curve and an underestimated amplitude of the synchronous pdf. This is a consequence of the discrete pdf oscillating around the continuous pdf due to inadequate intra-bin interpolation, so the minimum length $\hat L Y$ will correspond to the discrete pdf with a slightly smaller amplitude rather than the correct amplitude and discrete oscillations.

A closer look at Fig.~\ref{fig:observed} might raise some eyebrows\rev{. The lower-right panel provides a likely example of one of the possible artifacts of the method. The extracted signal exhibits dips between phases 0.0 and 0.1 and between phases 0.9 and 1.0 which are unlikely to be real and are more probably caused by} (1) the tolerance $\eta$ set too high so that convergence is not yet complete; (2) the signal-to-noise ratio of observations being too low; \rev{(3) the solution corresponds to a local minimum that it cannot get out of; }or (4) that the length of the asynchronous part of the overall curve does not constrain the smoothness of the synchronous pdf adequately well. While any combination of these might be at play, the reason \rev{for the deviations in Fig.~\ref{fig:observed} is the latter: the coherence of the 37$^\mathrm{th}$ harmonic has a high spectral power that prevents \emph{collective} excursions of bins to happen and further improve the smoothness of the synchronous signal}. It is certainly possible to generalize the cost function to depend on the length of the asynchronous part of the curve \emph{and} on the smoothness of the synchronous pdf, for example by tracing the relationship between the synchronous pdf length and the asynchronous curve length via the partial derivative $\partial \hat L \widetilde Y_k/\partial \hat L X_k$\rev{, but doing this would introduce coupling parameters that would adversely affect the non-parametric nature of the pdf. We also note that the example depicted in Fig.~\ref{fig:observed} is particularly pathological and, as such, it requires special treatment that the majority of other signals do not need.}

Finally, note that the \emph{levels} of the disentangled parts are completely arbitrary. The algorithm is using euclidean distance to determine the length of the residual curve, so any and all level information is subtracted out. Thus, we cannot tell from the algorithm output alone what is the mean of the pdf nor what is the mean of the asynchronous part of the light curve. This might be inferred from modeling considerations (i.e.~expected depths of eclipses from measured luminosities of the stars), but such information would need to be applied a-posteriori to the algorithm output.

\section{Discussion} \label{sec:conclusion}

It is tempting to take the synchronous pdf to represent the astrophysical signal of interest itself. That would be incorrect; instead, it should be taken as a \emph{generating function} for the observations of that signal. In other words, if realistic observational noise (say, a Poisson shot noise) were added to the synchronous pdf, that would be a fair representation of what we might observe with an ideal instrument, i.e.~an instrument that does not modify or modulate observations. The question, then, is what we mean by ``realistic'' when we say ``realistic observational noise'' in the context of the derived pdf. As the (stochastic, non-instrumental) noise inherent in the observations from which the pdf is derived fundamentally limits the accuracy of the pdf, it is precisely that level of noise that should be added to the generating pdf. The disentangled signal, thus, is \emph{not} the pdf itself, but the pdf \emph{plus} the representative stochastic noise of the original dataset.

The algorithm can be readily applied to vast numbers of strictly periodic variables, ranging from eclipsing binaries and transiting extra-solar planets to single-modal pulsating stars. There are two principal uses for the algorithm: (1) to detrend the data and derive a generating function for the signal of interest, and (2) to determine the instrumental response, coupled with noise properties. The first use is likely intuitive to anyone working with data, while the second use might not be immediately obvious. For example, we could run the algorithm on all periodic variables observed by \textsl{Kepler} to derive the asynchronous part of their light curves and then study the dependence of that residual signal as a function of CCD position, magnitude, time, etc. Having those ``anchor'' signals, the entire CCD can be mapped and all remaining, non-periodic objects can be detrended using 2-D interpolation. 

The algorithm can further be generalized to detrend \emph{any} signal that can be phase-folded. For example, including ETVs or orbital motion of the instrument due to barycentric motion, or any other effect that modulates the strictly periodic signal in a predictable manner allows us to phase-fold the data using a single period. Finally, multi-modal signals can also be detrended using the algorithm if the synchronous signal $X(t)$ in Eq.~(\ref{eq:signal}) is generalized as $\sum_i X_i(t)$, where each $X_i$ has a corresponding \emph{constant, independent} period (i.e.~it is \emph{not} a harmonic of any other input period or a combination of periods).

In this paper we deliberately refrained from optimizing and fine-tuning the algorithm to retain the clarity and to allow researchers to optimize it to their own needs that may differ from the general discussion presented above. In the online Appendix we provide complete scripts, written in \rev{P}ython 3 \citep{python1995}, to run the algorithm in jupyter notebooks and recreate the results depicted in Figs.~\ref{fig:synthetic} and \ref{fig:observed}. The notebooks and the data are also available at \texttt{http://keplerEBs.villanova.edu/DPS}.

\rev{A slightly more sophisticated, turn-key version of the code is released as open source, available from github\footnote{\texttt{https://github.com/aprsa/dips}} or via pip\footnote{\texttt{pip3 install dips}}. The released code includes three generalizations:}
\begin{description}

\item[Support for multi-processing] \rev{the computation of the slopes $\partial \hat L \widetilde Y_k/\partial X_i(\Phi_k)$ per Eq.~(\ref{eq:slopes}) is the bottleneck of the algorithm and can become a significant time sink. At the same time, the computation of the slope per bin is entirely independent and can be distributed to parallel processors when available. We use Python's \texttt{multiprocessing} module to achieve that. The computation of slopes is distributed across all available cores.}

\item[Slope-based convergence criterion] \rev{the basic algorithm presented here asserts that convergence is reached when the length change of the asynchronous curve dips below a certain threshold. This criterion is ad-hoc because it does not relate the length change and topology of the parameter space in any way. An alternative convergence criterion is to trace the magnitude of the slopes themselves and to assert convergence when the magnitude of the average down-step, i.e.~the product $\xi \left \langle \| \partial \hat L \widetilde Y_k/\partial X_i(\Phi_k) \| \right \rangle$, drops below a certain threshold. We \emph{start} with a predefined value of $\xi$ (say, $10^{-3}$) and reduce its value by an attenuation factor (say, 10\%) whenever the down-step taken was too large, i.e.~when the new length of the asynchronous curve is larger than the previous length of the asynchronous curve. Thus, as the algorithm approaches a minimum, the topology flattens and both $\xi$ and the mean absolute magnitude of the slopes tend to $0$. By connecting the tolerance for convergence with the step size, we directly relate the length of the residual curve to the topology of the parameter space.}

\item[Variable gradient descent] \rev{as stated above, the algorithm converges to \emph{a} minimum, but not necessarily to \emph{the} minimum. To avoid getting stuck in a local minimum, the algorithm does not descend along the gradient itself but, instead, along the randomly drawn gradient from the normal distribution centered on the actual gradient. The width of the distribution is a provided parameter and a reasonable value is between 5\% and 50\%. This is also known as the ``drunken sailor'' gradient descent.}

\end{description}

\rev{These additions to the algorithm make it more robust because the initial value of $\xi$ no longer affects convergence properties, tolerance is based on the shape of the parameter space and ``jittering'' the convergence path may help the algorithm avoid local minima. Of course, this cannot possibly satisfy all use cases out there so further adaptations will likely remain necessary by the users. With the source code available on github, we hope to make that task more tractable for everyone.}

\acknowledgements

The authors gratefully acknowledge support from the NSF grant \#1517460 and from the NASA grant 17-ADAP17-68 which led to the development of the presented algorithm.

\bibliography{main}

\end{document}